\documentclass[letterpaper,prl,twocolumn]{revtex4}
\usepackage{graphicx,bm,color}    

\begin{document}
\title{Zero-bias anomaly induced by the point defect in graphene}
\author{Wen-Min Huang$^1$, Jian-Ming Tang$^2$ and Hsiu-Hau Lin$^{1,3}$}
\date{\today}
\affiliation{$^1$ Department of Physics, National Tsing-Hua
university, Hsinchu 300, Taiwan\\
$^2$Department of Physics, University of New Hampshire, Durham, New Hampshire 03824-3520, USA\\
$^{3}$ Physics Division, National Center for Theoretical Sciences, Hsinchu 300, Taiwan}

\begin{abstract}
It is generally believed that a point defect in graphene gives rise to an impurity state at zero energy and causes a sharp peak in the local density of states near the defect site. We revisit the defect problem in graphene and find the general consensus incorrect. By both analytic and numeric methods, we show that the contribution to the local density of states from the impurity state vanishes in the thermodynamic limit. Instead, the pronounced peak of the zero-bias anomaly is a power-law singularity $1/|E|$ from infinite resonant peaks in the low-energy regime induced by the defect. Our finding shows that the peak shall be viewed as a collective phenomenon rather than a single impurity state in previous studies.
\end{abstract}
\maketitle


Graphene, a single-layer graphite composed of carbon atoms arranged in two-dimensional honeycomb lattice, is recently fabricated in laboratory and attracts intense attentions from both experimental and theoretical aspects\cite{Novo1,Novo2,Zhang,Zhou06}. One of the striking features of graphene is its linear dispersion, allowing a relativistic description by a pair of massless Dirac fermions in the low-energy regime\cite{Semonoff,Haldane}. The relativistic spectrum gives rise to interesting phenomena such as half-integer quantum Hall effect\cite{Novo3}, Klein paradox\cite{Katsnelson06}, edge magnetism\cite{Son06,Hikihara03} and so on\cite{Tombros07,Bostwick07,Park08,Neto07}. In addition, it has been known for a long time that a point defect in the massless Dirac system induces peculiar quasi-localized state\cite{Dong98}. Recent theoretical investigations confirm its presence as a pronounced peak in the local density of states (LDOS)\cite{Pereira06}. Furthermore, the first-principles calculations seem to indicate that there exists quantized magnetic moment associated with each defect in graphene\cite{Lehtinen04,Vozmediano,Wehling}. Since a point defect can be created by chemisorption of hydrogen atoms onto graphene\cite{Lehtinen04,Yazyev,Yazyev08}, these calculations suggest an alternative approach to fabricate spin qubits\cite{Trauzettel07} with relatively easy efforts, if the predicted moments survive the quantum fluctuations not included in these studies.

While it is not yet settled whether the quantum fluctuations will destroy the magnetic moment near a defect, it is generally believed that the zero-bias anomaly in the LDOS corresponds to the impurity state from the point defect\cite{Pereira08}. We revisit this problem with both analytic and numeric methods and find the general consensus on this issue is incorrect. We start from graphene in nanotorus geometry and investigate how the system evolves as the size becomes infinite in the thermodynamic limit. At finite system size $N$, the defect indeed gives rise to a zero-energy peak in LDOS as expected\cite{Lieb}. However, its spectral weight decays as $1/\ln N$ or $1/N$ (depending on whether the nanotorus is semiconducting or metallic) and eventually vanishes. This conclusion, overlooked by previous studies, is indeed reasonable since the impurity state is quasi-localized and should have no contribution to LDOS as the system size goes infinite.

How can we explain the zero-bias anomaly found in previous studies then? This paradox can be explained in two steps. First of all, our numerics show that the defect in graphene induces enormous resonant peaks in the LDOS at energies close to zero. Then, as the system size grows to infinity, these peaks crowd into zero energy and become singular. Both numerical and analytic approaches give $1/|E|$ power-law singularity with weak logarithmic corrections. That is to say, the zero-bias anomaly in the LDOS is not from a single impurity state. Instead, it is a power-law singularity from collective resonance induced by a single defect. This is remarkable that the impurity state in graphene dissolve into a (weaker) power-law singularity as the single-particle state disappears in one-dimensional interacting electron gas. It is suggesting that graphene is already at criticality so that introduction of a point defect reshuffles the LDOS leading to a power-law singularity rather than a well-defined delta-function (or broadened Lorentzian) peak. Our finding here has a significant impact on the idea of fabricating spin qubits by defects in graphene. Since the peak in the LDOS is in fact a power-law singularity from many resonant states, the quantum coherence between these states with realistic interactions will be tough to maintain for a working qubit.


Now we walk through the details which lead to the conclusions sketched in above. Because the band structure of graphene obtained by the first-principles calculations is well approximated by the nearest-neighbor hopping for the active $\pi$ orbitals\cite{Reich,Loiseau07}, it is sufficient to start from a tight-binding Hamiltonian and add a single defect at the origin,
\begin{equation}\label{Hamiltonian}
H=-t\sum_{\langle{\bf r},{\bf r'}\rangle}\left[c^{\dag}({\bf r})c({\bf r'})+c^{\dag}({\bf r'})c({\bf r}) \right] + V_0c^{\dag}({\bf 0})c({\bf 0}) \;,
\end{equation}
where $t$ is the nearest-neighbor hopping amplitude, and $V_0$ is the strength of the impurity potential. In the remaining part of the calculations, we mainly focus on the unitary limit $V_0 \to \infty$. Though the impurity state in the unitary limit has been solved analytically in a recent paper\cite{Pereira06}, it is insightful to rederive it with cares so that the evolution of the impurity state with the system size $N$ is clarified. We follow the definitions used in Ref. \cite{Pereira06} and reformulate the problem in more compact notations.

Consider a zigzag carbon nanotube with $N$ unit cells around the transverse direction. The total number of localized states at the left (right) of the defect is denoted as $N_L (N_R)$. Construct a $(N_L+N_R)$-dimensional vector to represent the solution,
$
\bm{A} = [a^{(L)}_{k_m}, (1+e^{ik_{m'}}) a^{(R)}_{k_{m'}}],
$
Meanwhile, we can also introduce $(N-1)$ vectors in the same space
$
\bm{B}_{j} = [e^{ik_m j}, e^{i k_{m'}j}],
$
where $j =1, 2, ..., N-1$. The boundary conditions\cite{Pereira06} then take the simple form $\bm{A} \cdot \bm{B}_j = 0$, i.e. we are looking for all linearly independent vectors $\bm{A}$ which is orthogonal to the subspace expanded by the $N-1$ vectors $\bm{B}_j$.

For semiconducting nanotubes ($N \neq 3n$), the quantized momentum does not cut through the Dirac points at $k=2\pi/3, 4\pi/3$. Thus, counting all localized states within $2\pi/3 < k_m < 4\pi/3$ gives $N_L = [(N+1)/3]$, where $[x]$ denotes the Gauss symbol. Similarly, it is straightforward to obtain $N_R = [(2N+1)/3]$. Since the combination of $k_m$ and $k_{m'}$ exhausts all quantized momenta in the Brillouin zone, $N_L+N_R=N$. The boundary conditions nail down the symmetric solution,
\begin{eqnarray}
\bm{A} = [1,1,...,1].
\label{original}
\end{eqnarray}
It is easy to verify that
$
\bm{A} \cdot \bm{B}_j = \sum_{n=1}^{N} e^{i2\pi n j/N} = N \delta_{j,0} =0,
$
satisfying all orthogonal criteria. Furthermore, because $\bm{B}_j$ expand a subspace of dimension $N-1$ in the $N$ dimensional space, only one solution is allowed. Normalization of the above solution involves a summation over all localized states, $\sum_{m}[1-2 \cos(k_m/2)] \sim \int dk/k \sim \ln N$. As a result, the spectral weight at $E=0$ scales as $1/\ln N$ in the thermodynamic limit.

\begin{figure}
\begin{center}
\includegraphics[width=7cm]{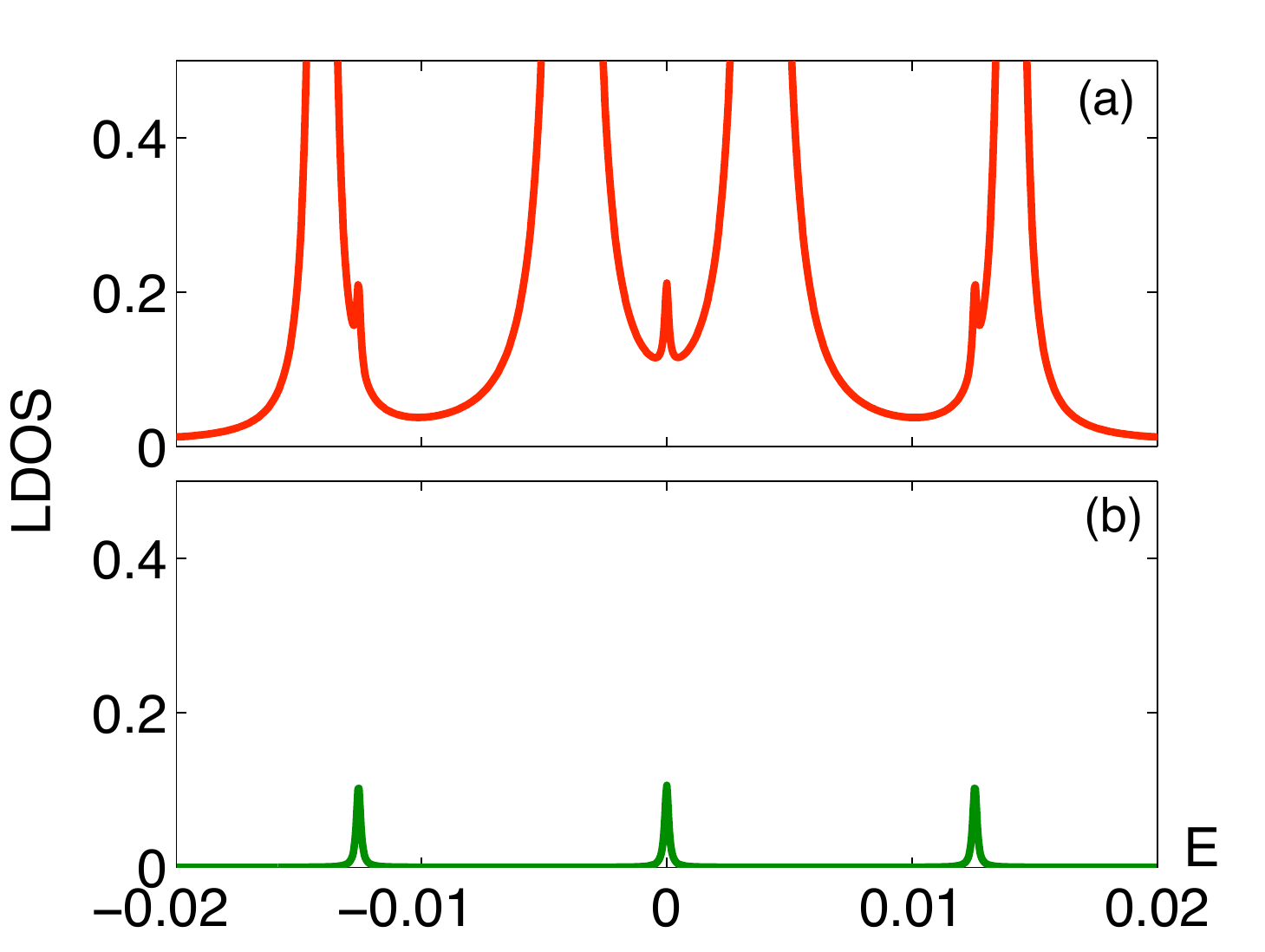}
\caption{(Color online) The LDOS for a nanotorus of the size $N =300\times 144$ (a) with and (b) without a point defect evaluated at its nearest neighbor site.}
\label{ResonantPeak}
\end{center}
\end{figure}

For metallic nanotubes ($N=3n$), there are two additional extended states at the Dirac points with the nodal structure. These extended states can hybridize with the localized states and should be included on both sides. Thus, $N_L = (N/3)+1$ and $N_R = (2N/3)+1$. The dimension of the vector space is larger $ N_L+N_R=N+2$ here. Let us count the number of independent solutions first. Subtracting the dimension of the vector space $N+2$ by the dimension $N-1$ expanded by $\bm{B}_j$, we expect {\em three} independent solutions. Let us start with the simple ones which do not involve the localized states,
\begin{eqnarray}
\bm{A}_{D1} &=& [\:\: {\it 1},{\it 0},\:\: 0,...,0,\:\: {\it -1},{\it 0},\:\: 0,...,0\:\:],
\nonumber\\
\bm{A}_{D2} &=& [\:\: {\it 0},{\it 1},\:\: 0,...,0,\:\: {\it 0},{\it -1},\:\: 0,...,0\:\: ].
\end{eqnarray}
For clarity, the components corresponding to the Dirac points are in italics while those for localized states are in roman. The orthogonal criteria are trivially satisfied. In fact, these two solutions are just the extended states at the Dirac points. The third solution is
\begin{eqnarray}
\bm{A}_{m} = \left[\:\: {\it \frac{1}{2}},{\it \frac{1}{2}},\:\: 1,...,1,\:\: {\it \frac{1}{2}},{\it \frac{1}{2}},\:\:1,...,1 \right].
\label{correct}
\end{eqnarray}
Although the solution is similar to that in Eq.~(\ref{original}), its hybridization with the extended state at the Dirac points makes it an extended state as well. Thus, despite the superficial resemblance, its contribution to the spectral weight at $E=0$ is dramatically different and scales down much faster as $1/N$.


\begin{figure}
\begin{center}
\includegraphics[width=7.3cm]{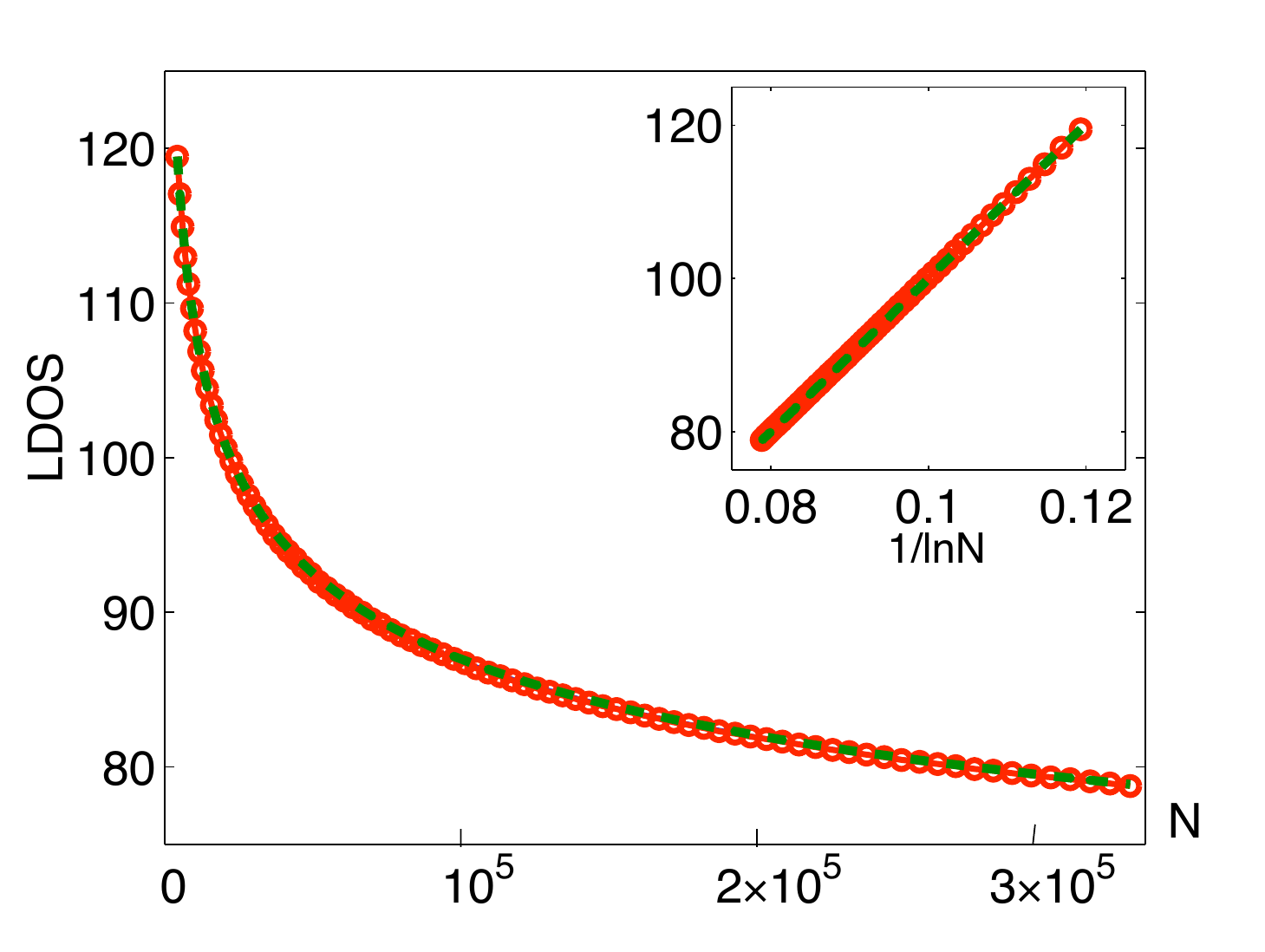}
\caption{\label{SidePeak}
(Color online) The height of an individual resonant peak in the LDOS close to the zero energy versus the system size $N$. The green dashed line is the fit to the $1/\ln N$ scaling.}
\end{center}
\end{figure}

The above analytic solutions exclude the impurity state as the cause for the zero-bias anomaly of the LDOS in graphene. We also perform numerical diagonalization to confirm the different trends for semiconducting and metallic cases. Note that , in addition to the impurity state, the defect in graphene also generates resonant states at energies close to $E=0$ inaccessible by analytic approach. We employ numerical diagonalization of the carbon nanotorus (by jointing the two ends of the zigzag carbon nanotube together) to address this issue. For comparison, we computed the LDOS for a metallic nanotorus with and without a point defect in Fig.~\ref{ResonantPeak}. The change of the LDOS {\em right at} zero energy caused by the defect is almost invisible. But, the induced resonant peaks {\em close to} the zero energy are enormous and largely change the profile of the LDOS. It is reasonable to expect that the resonant peaks are far more important than the impurity state at zero energy in the thermodynamic limit.

Furthermore, our numerics also reveal that the height of each individual resonant peak at finite (but small) energy scales as $1/\ln N$ with the system size as shown in Fig.~\ref{SidePeak}. This $1/\ln N$ dependence of the resonant peaks bridge a gap between the semiconducting and the metallic nanotori in thermodynamic limit. Note that the impurity state by itself in the semiconducting or in the metallic nanotorus gives $1/\ln N$ or $1/N$ contribution to the zero-energy LDOS. The difference in LDOS due to the impurity state in the two classes is huge in  a large but finite system. The merger of the $1/\ln N$-resonant peaks resolves this discrepancy between the semiconducting and metallic systems and establishes the smooth thermodynamic limit to the two-dimensional graphene.

\begin{figure}
\begin{center}
\includegraphics[width=7.3cm]{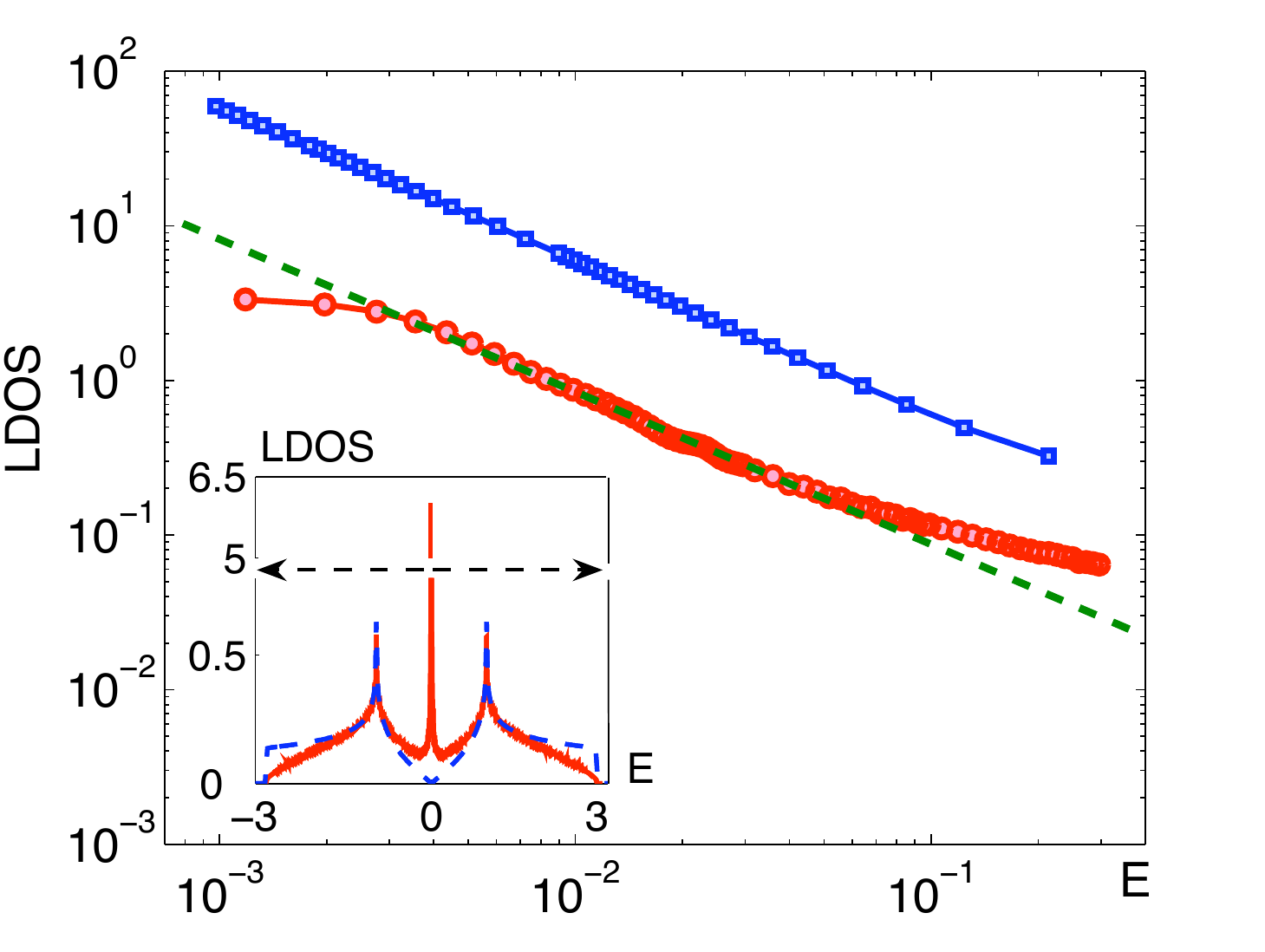}
\caption{\label{Scaling}
(Color online) The LDOS at the nearest-neighbor site of the defect for a nanotorus of the size $N=564\times564$ with a broadening factor $0.004$ (red circles) and in the two dimensional limit with a broadening factor $10^{-5}$ (blue squares). The green dashed line of $1/|E|$ is the guide to the eyes. In the inset, the LDOS at the nearest-neighbor site (shown in red solid line) is compared to that deep inside the bulk (shown in blue dashed line). Since the peak at zero energy is high, it is chopped off by the double-arrow dashed line for clarity.
}
\end{center}
\end{figure}

The remaining task is to understand how these resonant peaks crowd into the zero-energy regime as the system size goes to infinity. The numerical results are shown in Fig.~\ref{Scaling}. In the inset, we compare the LDOS near the defect and in the bulk. As in previous studies, the most noticeable difference is the sharp peak at $E=0$. However, one may also observe that there exists large spectral weight transfer from the higher-energy regime to the lower energy. This is an indirect hint that the sharp peak at zero energy may not attribute to a single impurity state. To illustrate this point, the LDOS shown in the log-log plot reveals the $1/|E|$ power-law dependence. The deviation from the power-law singularity at extremely small energy where it is rounded off comes from the Lorentzian energy broadening factor introduced in the numerical calculations.


To further strengthen our numerical findings, we solve the Dyson equation for graphene in the presence of a point defect. The change of the LDOS can be expressed in terms of the non-interacting retarded Green's functions, $G^{\pm}_{0}({\bf k},E)=1/(E\pm|h({\bf k})|+i\eta)$, representing the propagation of particles and holes with the dispersion $h({\bf k})=t+2t\cos(k_x/2)e^{i\sqrt{3}k_y/2}$ from the tight-binding model. Following standard Green's function techniques, the change of the LDOS at the first-nearest-neighbor sites (B) of the defect (located at the A site) is
\begin{eqnarray}\label{deltarho1}
\Delta\rho(E)&=&-\frac{1}{\pi}{\rm Im}\left[\frac{V_0G_{AB}G_{BA}}{1-V_0G_{AA}}\right],
\end{eqnarray}
where $G_{\Lambda \Lambda'}$ with $\Lambda, \Lambda' = A, B$ are the Green's functions between sites $\Lambda$ and $\Lambda'$. In the unitary limit $V_0 \to \infty$, the change of the LDOS no longer depends on the strength of the impurity potential as expected. 

For graphene, it is straightforward to show that symmetries between these Green's functions lead to the relations, $G_{AA}=G_{BB}$ and $G_{AB}=G_{BA}$,
\begin{eqnarray}
G_{AA}(E) &=&\int \frac{d^2{\bf k}}{8\pi^2}[G^-_{0}({\bf k},E)+G^+_{0}({\bf k},E)],
\\
G_{AB}(E) &=& \int \frac{d^2{\bf k}}{8\pi^2}\frac{h({\bf k})}{|h({\bf k})|}[G^-_{0}({\bf k},E)-G^+_{0}({\bf k},E)].
\end{eqnarray}
These Green's functions can be solved numerically to compute the LDOS. The results are shown as blue squares in Fig.~\ref{Scaling}. It delivers the same power-law singularity in the LDOS. It is also interesting to notice that, though with the same exponent, the absolute values of the LDOS for the large nanotorus with $N = 564 \times 564$ and the two-dimensional graphene are different -- another hint for the collective phenomena rather than a single impurity state for the zero-bias anomaly.

The power-law singularity can be derived from the linear dispersion near the Dirac cones. Carrying out the angular part of the integral near the Dirac points with linear dispersion, the Green's functions are approximately
\begin{eqnarray}
G_{AA}(E) &\sim& E\ln (E^2/\Lambda^2)+i |E|,
\\
G_{AB}(E) &\sim& \Lambda^2-E^2\ln (E^2/\Lambda^2)+i\: {\rm sign}(E) E^2,
\end{eqnarray}
where $\Lambda$ is a momentum cutoff introduced for linearizing the spectrum. Since $G_{AB}(E)$ appears in the numerator, it can be treated as a constant in low-energy limit. Therefore, the change of the LDOS is $\Delta\rho(E) \sim 
{\rm Im}[G_{AA}]/({\rm Re}[G_{AA}]^2+{\rm Im}[G_{AA}]^2)$ and the singularity near $E=0$ emerges,
\begin{eqnarray}\label{deltarho2}
\Delta\rho(E) \sim \frac{1}{|E|\left({\rm ln|E|}\right)^2}.
\end{eqnarray}
Because the logarithmic correction is very weak (beyond the resolution of our numerical results), the singularity is essentially a power-law $1/|E|$ and agrees with our previous numerical findings.

The above calculations are readily generalized to other systems with gapless dispersion $h({\bf k}) \sim |{\bf k}|^{\alpha}$. Following the same steps, it is straightforward to show that the imaginary of the Green's function follows the energy dependence, ${\rm Im}[G_{AA}] \sim |E|^{(2/\alpha)-1}$. Ignoring the logarithmic correction, the real part of the Green's function shares the same energy dependence. Thus, the change of the LDOS exhibits the power-law $\Delta\rho(E)\propto |E|^{1-(2/\alpha)}$. For double-layer graphene, the dispersion is quadratic with $\alpha=2$, giving a constant $\Delta \rho(E)$ at zero energy. Our calculations thus predict that the power-law singularity is absent for the bilayer graphene in sharp contrast to the $1/|E|$ singularity for the single-layer.

So far, no correlation effect is taken into account. Though it is generally believed that the mutual interactions between electrons are likely to be irrelevant in graphene, there are evidences that the correlation effects may be significant near defects or open boundaries. Further in-depth investigations are necessary to explore how the power-law singularity evolves with the inclusion of electronic correlations. However, it is worth mentioning how the exponent changes in the analogous one-dimensional interacting system. Upon the inclusion of interaction, the delta-function for the quasi-particle changes into $1/|E|$ power-law singularity. Increasing the strength of the interaction suppresses the density of states near the zero energy and gradually changes the exponent from the negative $\gamma=-1$ to positive values, exhibiting the so-called ``pseudo-gap" behavior. We may expect similar trend happens for defect in graphene but it needs further investigations.

In conclusions, we revisit the defect problem in graphene and find the sharp peak in the LDOS is not due to the impurity state at zero energy. Instead, it is a power-law singularity $1/|E|$ from infinite resonant peaks in the low-energy regime induced by the defect. Our finding here has a significant impact on the idea of fabricating spin qubits by defects in graphene. Since the peak in the LDOS is in fact a power-law singularity from many resonant states, the quantum coherence between these states with realistic interactions will be tough to maintain for a working qubit.

We acknowledge supports from the National Science Council in Taiwan through grants NSC-96-2112-M-007-004 and NSC-97-2112-M-007-022-MY3. Financial supports and friendly environment provided by the National Center for Theoretical Sciences in Taiwan are also greatly appreciated.

\end{document}